\documentclass[aps,twocolumn,superscriptaddress]{revtex4}
\usepackage{graphicx,subfigure}
\usepackage{amsthm}
\usepackage{amsmath,stmaryrd}
\usepackage{color,colordvi}
\usepackage[T1]{fontenc}
\usepackage[utf8]{inputenc}
\usepackage{lmodern}

\begin{document}
\bibliographystyle{revtex4}
\draft


\title{On electron heating in a low pressure  capacitively coupled oxygen discharge}



\author{J. T. Gudmundsson}
\email[]{tumi@hi.is}

\affiliation{Science Institute, University of Iceland,
                Dunhaga 3, IS-107 Reykjavik, Iceland}

\affiliation{Department of Space and Plasma Physics, School of Electrical Engineering, 
KTH--Royal Institute of Technology, SE-100 44, Stockholm, Sweden}

\author{D. I. Snorrason}

\affiliation{Science Institute, University of Iceland,
                Dunhaga 3, IS-107 Reykjavik, Iceland}

\date{\today}

\begin{abstract}
We use the one-dimensional object-oriented particle-in-cell Monte Carlo collision code {\tt oopd1}
to explore the charged particle densities, the electronegativity,
the  electron energy probability function
(EEPF), and the  electron heating mechanism  in  a single frequency 
 capacitively coupled oxygen discharge when the applied voltage amplitude is varied. We explore
discharges operated at 10 mTorr, where electron heating within the plasma bulk (the electronegative core) 
dominates, and  at 50 mTorr where sheath heating dominates.  At 10 mTorr the discharge 
is operated in combined drift-ambipolar (DA)  and $\alpha$-mode
and at 50 mTorr it is operated in pure  $\alpha$-mode.
At 10 mTorr the effective electron temperature is high and increases with increased
driving voltage amplitude, while at 50 mTorr the effective electron temperature is much lower,
in particular within the electronegative core, where it is roughly 0.2 -- 0.3 eV, and varies only a little with the 
voltage amplitude.
\end{abstract}

\maketitle


\section{Introduction}
\label{intro}
Low temperature low pressure radio frequency (rf) driven capacitively coupled plasma (CCP) 
discharges have been used for several decades for etching and deposition of thin films.
The CCP consist of two parallel electrodes, typically of radius of a few tens of cm, separated by a few cm, 
and biased by a radio-frequency power supply, often operated at 13.56 MHz.
In the CCP a  plasma forms between the electrodes, from which it is separated by space charge sheaths. 
Nearly all the applied voltage appears across the oscillating sheaths.
In a single-frequency CCP the ion flux ($\propto$ plasma density)
and ion energy ($\propto$ sheath voltage) cannot be varied independently.
The plasma parameters such as the electron density, the ion densities, the  effective electron temperature, 
and the electron energy distribution depend on the operating condition of the discharge including the
gas composition, gas pressure, applied voltage, reactor geometry and the electrode material.  
These operating parameters
dictate the mechanisms by which power is transferred to the electrons for sustaining the discharge.  
When operated at low neutral gas pressure
collisionless heating mechanism effectively transfers energy to the electrons. This
is due to a rapid movement of the electrode sheaths that contributes to the electron heating
via  stochastic or collisionless heating by the expanding sheaths \cite{lieberman98:955,gozadinos01:117}.  
As the electrons  interact with the moving sheaths, they can
be either cooled (collapsing sheath) or heated (expanding sheath).
When an energetic electron bounces back and forth between the two 
sheaths and hits each sheath during its expansion phase, it will
be heated multiple times. This effect is referred to as  electron bounce
resonance heating (BRH) \cite{liu11:055002}.
 The sheath motion and thus the stochastic heating can also be enhanced by 
self-excited non-linear plasma
series resonance (PSR) oscillations 
\cite{czarnetzki06:123503,donko09:131501,schungel15:044009,wilczek15:024002,wilczek16:063514}.
At higher pressures 
some of the  power is deposited by ohmic heating in the 
bulk plasma due to collisional momentum transfer
between the oscillating electrons and the neutrals.   For this mechanism to be dominant
the electron neutral mean free path must be smaller than or comparable to the discharge dimensions. 
When a discharge is sustained through collisional ohmic heating and/or
 stochastic heating it is commonly referred 
to as the $\alpha$-mode \cite{belenguer90:4447}.
At high applied voltages and pressures secondary electron emission can 
contribute or even dominate the ionization processes.
This  operation regime is referred to as $\gamma$-mode \cite{belenguer90:4447}.
In electronegative discharges large electron density gradients with local maxima 
of the electron density at the sheath edges, can develop within the rf period
and lead to generation of ambipolar fields, also  drift fields can arise due  to the low bulk plasma 
conductivity. These fields  can accelerate the electrons and are thus referred to
as drift-ambipolar (DA) mode \cite{schulze11:275001,derzsi15:346}.  

Oxygen discharges are widely used in plasma materials 
processing including oxidation of silicon \cite{pulfrey73:1529,kawai94:2223}, 
ashing of photoresist \cite{hartney89:1,kalpakjian94:1351}, 
and surface modification  of polymer films \cite{vesel17:293001}.
The properties of the oxygen discharge depend heavily on the  accumulation of the  singlet metastable 
oxygen  molecules, which are  known to play a significant role in the 
oxygen discharge  \cite{thompson61:519,katsch99:2023}.  
Earlier we have demonstrated that these singlet metastable molecular states influence the 
electron kinetics and the electron heating mechanism in the capacitively coupled oxygen 
discharge operated at a single frequency of 13.56 MHz 
\cite{gudmundsson15:035016,gudmundsson15:153302,hannesdottir16:055002,gudmundsson17:120001} 
as well as the ion energy  distribution in both single and dual frequency discharges
 \cite{hannesdottir17:175201}. We found that at low pressure (10 mTorr), 
electron heating in the bulk plasma (the electronegative core) dominates, while
 at higher pressures (50 -- 500 mTorr) 
the electron heating occurs mainly in the  sheath region. 
We demonstrated  that the detachment by the singlet molecular metastable states is the process that has the most 
influence on the electron heating process  in the higher pressure regime (50 mTorr), while it has only a small 
influence at lower pressure (10 mTorr) \cite{gudmundsson15:153302,hannesdottir16:055002,gudmundsson17:120001}.  
The presence of the singlet metastable molecules in the discharge model lowers the electronegativty and the
effective electron temperature 
and moves the electron heating from the plasma bulk to the sheath region when operated at 
50 -- 500 mTorr \cite{gudmundsson15:035016,gudmundsson15:153302}. 

In all of these studies we assumed the voltage amplitude to be fixed at  222 V.  
Here we study  how the applied voltage influences the charged particle densities,
 the electron heating processes, the electron energy probability function (EEPF) 
and the effective electron temperature in the single frequency voltage 
driven capacitively coupled oxygen discharge while 
we keep the  discharge pressure at either 10 or 50 mTorr.    
The simulation parameters and the cases explored are defined in section \ref{cases} and the results are discussed  
in section \ref{results}.  We discuss the electron heating rate, its spatio-temporal behavior and
the time averaged profile, the effective electron temperature and the particle densities and how these 
parameters depend on the driving voltage amplitude.  
We also discuss the center and average  electronegativity. 
Finally concluding remarks are given in section \ref{conclusion}.

\section{The simulation}
\label{cases}

The one-dimensional object-oriented particle-in-cell Monte Carlo collision (PIC/MCC) code {\tt oopd1}
\cite{hammel03:66,verboncoeur95:199} is here applied to study the capacitively coupled oxygen discharge.
In 1d-3v PIC codes  the model system has one spatial dimension and three velocity components.
Earlier we discussed the capability of the {\tt oopd1} code,  the advantages and improvements
compared to the  well established {\tt xpdp1} code, and benchmarked it against the {\tt xpdp1} 
code for a capacitively coupled discharge with a simplified oxygen discharge model \cite{gudmundsson13:035011}.   
The  {\tt xpdp1} code included only reaction set for oxygen molecules in the ground state O$_2$(X$^3\Sigma_g^-)$, O$_2^+$-ions,
O$^-$-ions and electrons and the neutral particles were not treated kinetically \cite{vahedi95:179}.  
Particle weight is the number of real particles  each superparticle represents, 
i.e.~the ratio of the number of
physical particles to computational particles. In {\tt oopd1} the various particles can have different weights.
As the neutral gas density  is much higher than the densities of charged species, different
weights allow us to treat both charged particles and neutral
particles kinetically.  
In our earlier work we added oxygen atoms in the ground state O$(^3$P) and ions of the oxygen 
atom O$^+$ and the relevant reactions to the {\tt oopd1} discharge model \cite{gudmundsson13:035011}. 
In subsequent studies we added the singlet  metastable molecule O$_2$(a$^1\Delta_g$) and
the metastable oxygen atom  O$(^1$D)  \cite{gudmundsson15:035016}, and the singlet
metastable molecule O$_2$(b$^1\Sigma$) \cite{hannesdottir16:055002}, to the  reaction set.
Furthermore, the discharge model now includes energy dependent secondary electron emission coefficients
for oxygen ions and neutrals  as they bombard the electrodes \cite{hannesdottir16:055002}.
For this current work the discharge model contains nine species: electrons, the ground state neutrals O($^3$P) 
and O$_2(\mathrm{X}^3\Sigma_g^-)$, the negative ion O$^-$, the positive ions O$^+$ and O$_2^+$, and the metastables 
O($^1$D), O$_2(\mathrm{a}^1\Delta_g)$ and O$_2$(b$^1\Sigma_g^+)$.
The full oxygen reaction set and the cross sections used have been discussed in our earlier works 
and will not be repeated here \cite{gudmundsson15:035016,hannesdottir16:055002,gudmundsson13:035011}.

We assume a  capacitively coupled discharge where one of the electrodes is driven by an rf voltage
\begin{equation}
V(t) = V_0 \sin( 2 \pi f t)
\end{equation}
while the other is grounded.  Here $V_0$ is the voltage amplitude, $f$ the driving frequency and, $t$ is the time. 
For this study we allow the applied voltage  amplitude $V_0$ to vary from 100 to 500 V while the  electrode
 separation is kept fixed at  4.5 cm, the driving frequency is 13.56 MHz, and a capacitor of 1 F is connected in series with the voltage source.
We assume discharges operated at pressures of 10 and 50 mTorr and assume geometrically symmetric electrodes.
  The discharge electrode separation is assumed to be small compared to electrode diameter so that the discharge can be
treated as one dimensional. 
The time step $\Delta t$ and the grid spacing $\Delta x$  resolve the electron plasma frequency and the
electron Debye length of the low-energy electrons, respectively, according to
$\omega_{\rm pe} \Delta t <  0.2$ where $\omega_{\rm pe}$ is the electron plasma frequency, and the simulation
grid is uniform and consists of 1000 cells. The electron time step is  $3.68 \times 10^{-11}$ s.  The simulation was run
for $5.5 \times 10^6$ time steps or 2750 rf cycles.   It takes roughly  1700 rf cycles to reach equilibrium 
for all particles and  the time averaged plasma parameters shown, such as the  densities, the electron heating rate, 
and the effective electron temperature, are averages over 1000 rf cycles. 
All particle interactions are treated by the Monte Carlo method with a null-collision scheme \cite{birdsall91:65}.
For the heavy particles we use a sub-cycling and the heavy particles are advanced every  16 electron time steps
 and we assume that the initial density profiles are parabolic  \cite{kawamura00:413}.

The kinetics of the charged  particles (electrons, O$^+_2$, O$^+$ and O$^-$-ions) was followed for all energies.
The neutral gas density is much higher than the densities of charged species, 
so that the neutral species at thermal energies (below a certain cut-off energy) 
are treated as a background with fixed
density and temperature and maintained uniformly in space. 
These neutral background species are assumed to have a
Maxwellian velocity distribution at the gas temperature (here~$T_{\mathrm{n}}$ = 26 meV).  
\begin{table*}
\caption{The parameters of the simulation, the particle weight, the threshold above 
which kinetics of the neutral particles are followed, the wall 
recombination and quenching coefficients used, and the partial pressure of the neutral 
species of the uniform background gas.}
 \label{simpar}
 \begin{tabular}{lllll}
\hline \hline
Species & particle weight & threshold  & wall quenching  & partial pressure  \\
        &                 &  [meV]    &  or recombination   &  [\%] \\
        &                 &           &  coefficients      &  \\
\hline
 O$_2(\mathrm{X}^3 \Sigma_g^-)$     &  $5 \times 10^9$  & 500  & 1.0 & 90.648              \\    
 O$_2$($\mathrm{a}^1\Delta_g$)      &  $5 \times 10^9$  & 100  &0.007  \cite{sharpless89:7947} & 4.4039   \\    
 O$_2$($\mathrm{b}^1\Sigma_g$)      &  $5 \times 10^7$  & 100  &0.1  & 4.4039   \\    
  O($^3$P)                          &  $5 \times 10^8$  & 500  &0.5 \cite{booth91:611}  &  0.519       \\    
  O($^1$D)                          &  $5 \times 10^8$  & 50   &1.0 & 0.0281 \\ 
      O$_2^+$                       &   $10^7$          & -  &     -  &  -       \\ 
      O$^+$                         &   $10^6$          & -  &   -   &   -     \\ 
      O$^-$                         &   $5 \times 10^6$ & -  &   -    &  -     \\ 
      e                            &  $1 \times 10^8$  & -  &   -    &  -     \\ 
\hline \hline         
 \end{tabular} 
\end{table*}
 We used a volume averaged (global) model \cite{thorsteinsson10:055008} to determine the partial pressure for each of
the thermal neutral species at 50 mTorr only and use these values at 10 mTorr as well.  These calculations give atomic
oxygen partial pressure of 0.519 \% which corresponds to atomic oxygen density of $8.3 \times 10^{18}$ m$^{-3}$ at 50 mTorr and 
  $1.6 \times 10^{18}$ m$^{-3}$ at 10 mTorr.  These are somewhat higher than the values of $0.5 - 3.2  \times 10^{18}$ m$^{-3}$, increasing with
increased applied power up to 200 W, reported by Kitajima et al.~\cite{kitajima04:2670} at 50 mTorr.  At 75 mTorr 
Katsch et al.~\cite{katsch00:6232} find the atomic density to be $ 5 \times 10^{18}$ m$^{-3}$ using TALIF, 
which is lower than $1.4 \times 10^{19}$ m$^{-3}$ if based on oxygen partial pressure of 0.519 \%.
 Also Kechkar et al.~\cite{kechkar13:045013,kechkar15t} measured the atomic oxygen density, by TALIF and actinometry in
a slightly asymmetric capacitively coupled discharge 
at 100 mTorr and 100 W, to be $1.4 \times 10^{20}$ m$^{-3}$, which is higher than  $1.9 \times 10^{19}$ m$^{-3}$ if based on 
oxygen partial pressure of 0.519 \%.  Thus this 
 atomic oxygen partial pressure can both be  over and under estimate. However, it is clear that
 at these densities the  role of atomic oxygen in these discharges is not very significant.
All the neutral species are treated kinetically as 
particles if their energy exceeds a preset threshold value.
The threshold values used here for the various neutral species are listed in Table \ref{simpar}.
As a neutral species hits the electrode it returns as a thermal particle with a given probability
and atoms can recombine to form a thermal molecule with the given probability.
Table \ref{simpar} lists all the wall recombination and quenching coefficients used 
for the neutral species here.  The wall recombination coefficient  for the neutral
atoms in ground state O($^3$P) is assumed to be 0.5 as measured by Booth and Sadeghi 
\cite{booth91:611} for a pure oxygen discharge in a stainless steel
reactor at 2 mTorr.  The choice of the  wall recombination coefficient for atomic oxygen has a 
significant influence on the atomic oxygen density and the type of electrode material 
 may explain the discrepancy in the experimental results discussed above.
As the metastable atom O($^1$D) hits the electrode we assume half of the atoms are
quenched forming O($^3$P) and the other half recombines to form the ground state molecule  O$_2(\mathrm{X}^3 \Sigma_g^-)$.
For O$_2$(a$^1\Delta_g$) we use a quenching probability of 0.007 estimated by Sharpless and Slanger 
 for iron while for aluminum they estimate the quenching probability to be $< 10^{-3}$ 
\cite{sharpless89:7947}.
A value of 0.006 is suggested by Derzsi et al.~\cite{derzsi16:015004}, found by comparing
PIC/MCC simulation with experimental findings which in their system (aluminium electrodes and $L = 2.5$ cm)  leads to 
[O$_2$(a$^1\Delta_{\rm g}$)]/[O$_2(\mathrm{X}^3\Sigma_g^-)$] density ratio that is on the order of 0.1.
 They point out that  there are significant changes in the  electronegativity as this parameter is 
varied.  Using a 1D fluid model Greb et al.~\cite{greb15:044003} demonstrated  that the electronegativity depends strongly 
on the  O$_2$(a$^1\Delta_{\rm g}$)  surface quenching probability. They
 argue that increased quenching coefficient leads to 
decreased  O$_2$(a$^1\Delta_{\rm g}$) density and thus decreased detachment by the   O$_2$(a$^1\Delta_{\rm g}$) 
state and thus higher negative ion density.    
We assume that the quenching coefficient for O$_2$(b$^1\Sigma_{\rm g}^+$)  at the electrodes  to be 0.1.  This 
assumption is based on the suggestion that the quenching coefficient  for the 
b$^1\Sigma_{\rm g}^+$ state is about two orders of magnitude larger than for the a$^1\Delta_{\rm g}$ state \cite{obrien70:3832}.  We neglect the reflection of electrons from  the electrodes.
The electrodes are assumed to be  identical,  made of stainless steel, 
and the surface coefficients are kept the same  at both 
electrodes.

\section{Results and Discussion}
\label{results}

\begin{figure}
\resizebox{0.45\textwidth}{!}{%
  \includegraphics{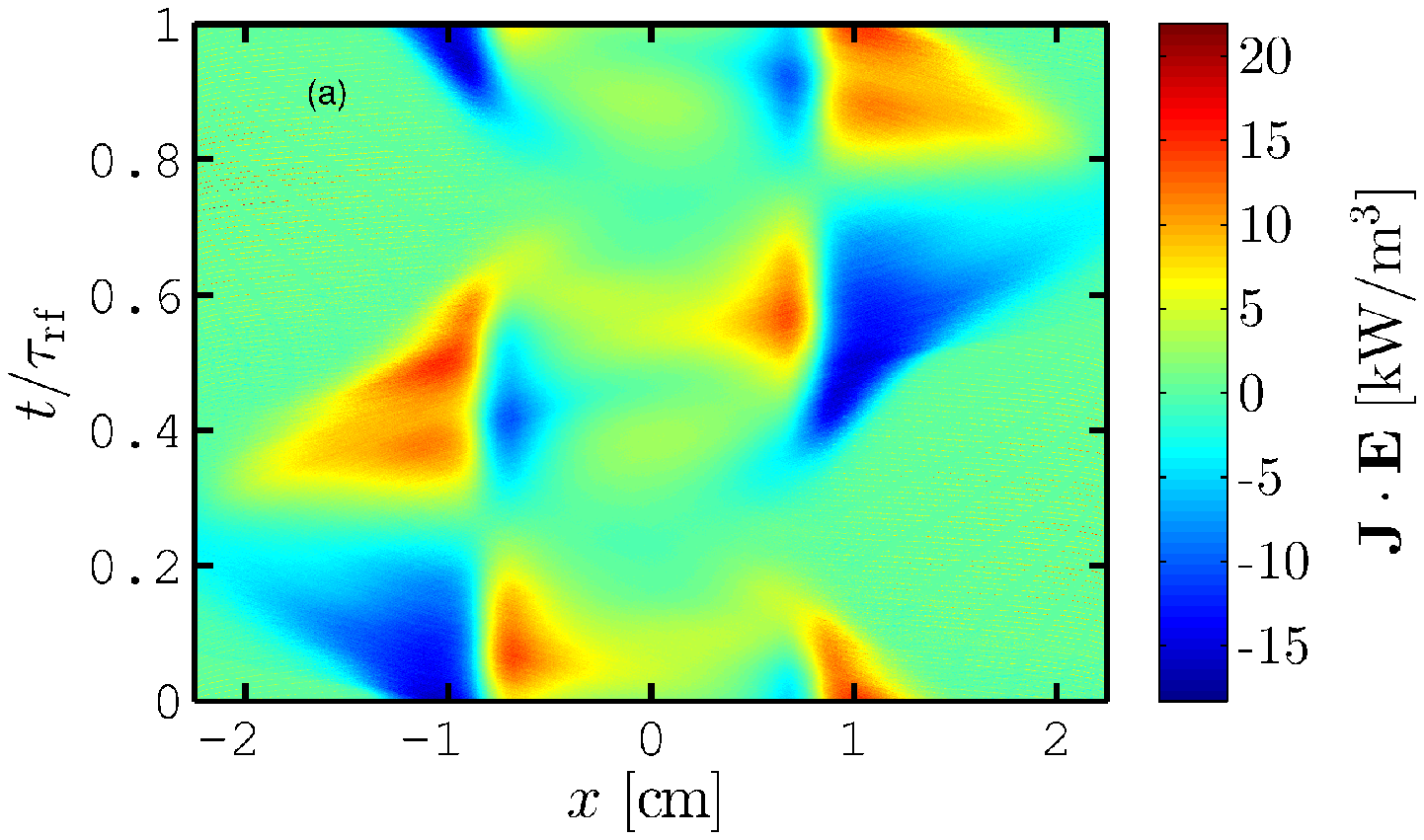}}
\resizebox{0.45\textwidth}{!}{%
  \includegraphics{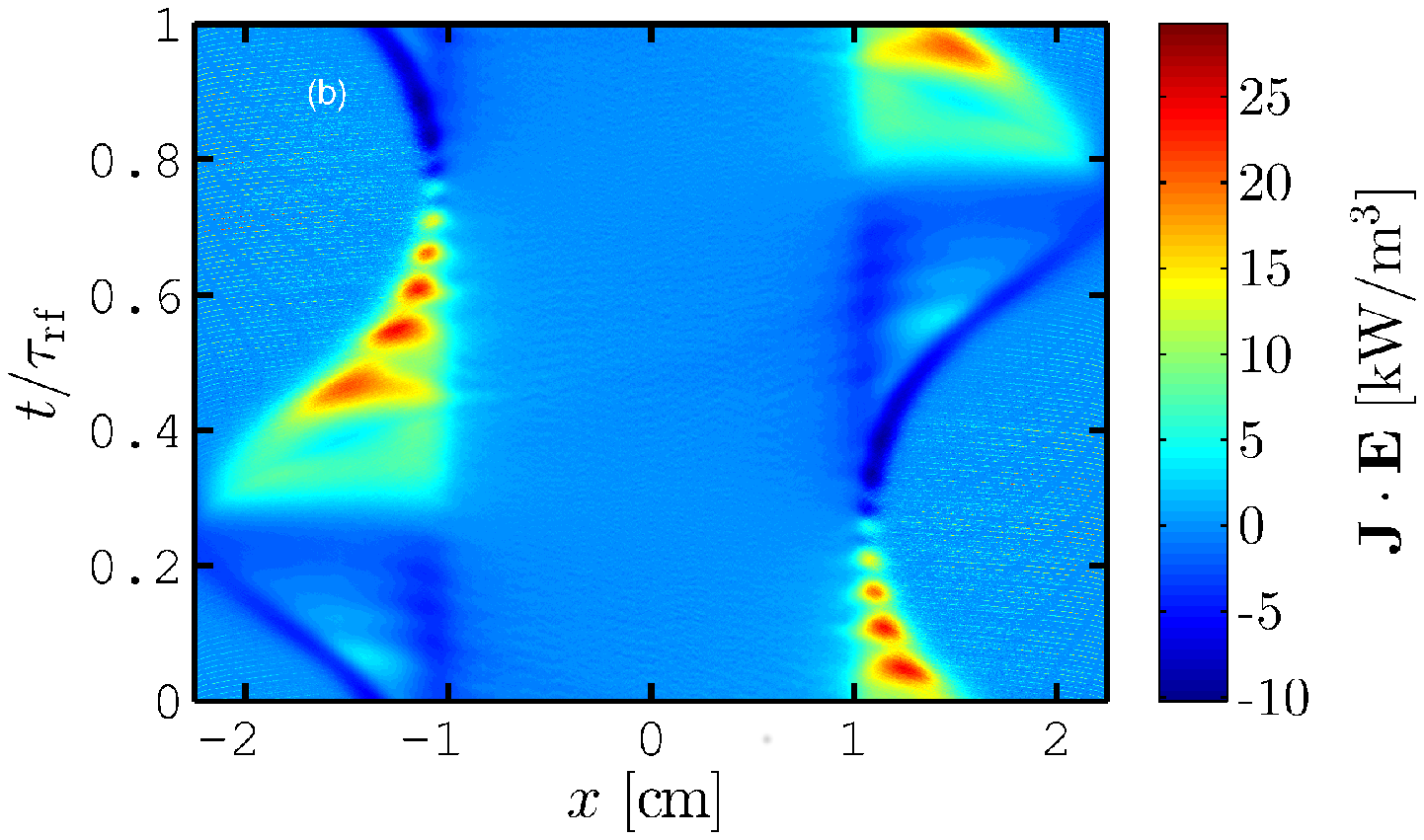}}
\caption{The spatio-temporal behavior of the  electron heating rate for a parallel 
plate capacitively coupled oxygen discharge with a gap
separation of 4.5 cm driven at 13.56 MHz  with $V_0 = 300$ V operated at (a) 10 mTorr, and (b)  50 mTorr.  }
\label{jdote}       
\end{figure}

\begin{figure}
\resizebox{0.45\textwidth}{!}{%
  \includegraphics{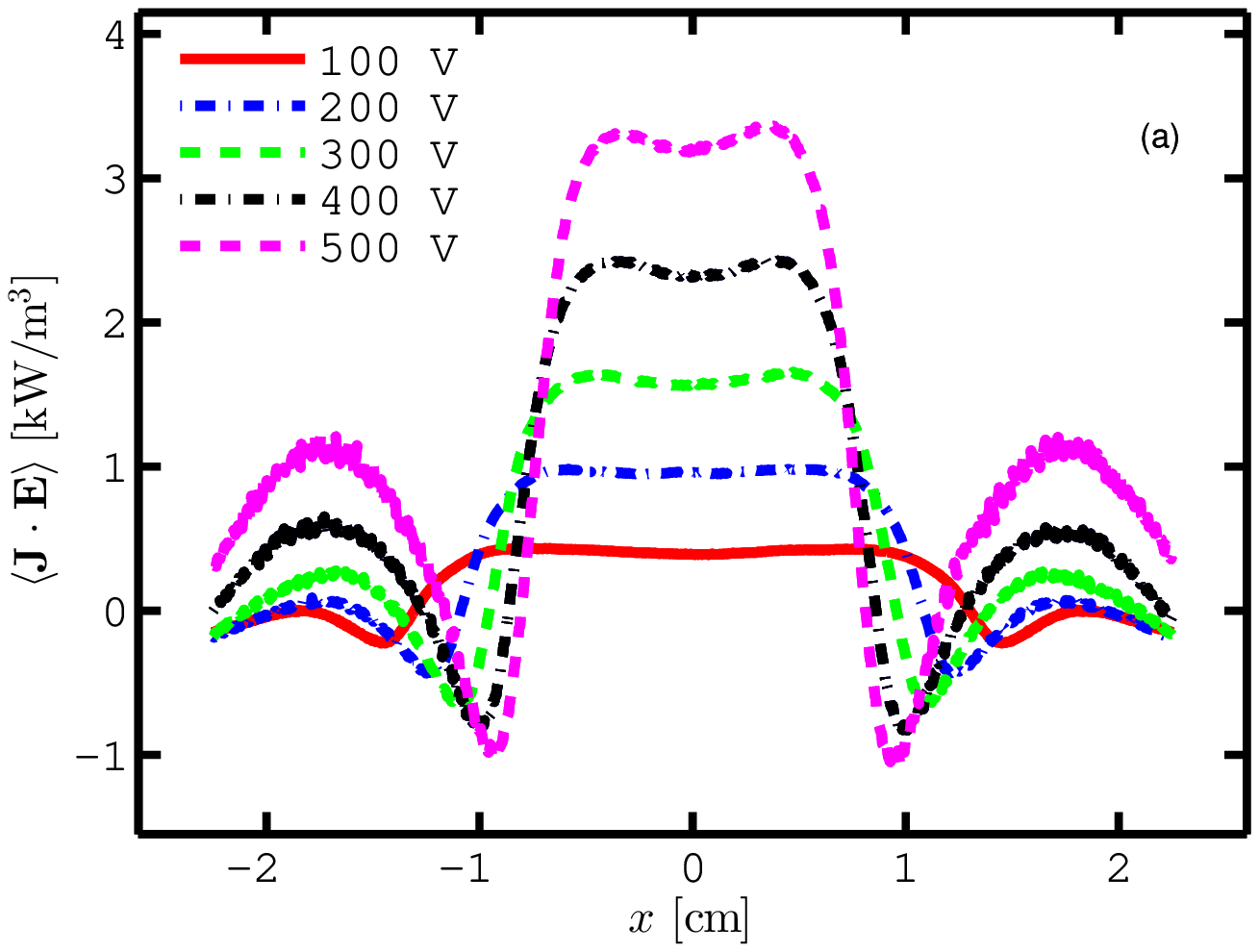}}
\resizebox{0.45\textwidth}{!}{%
  \includegraphics{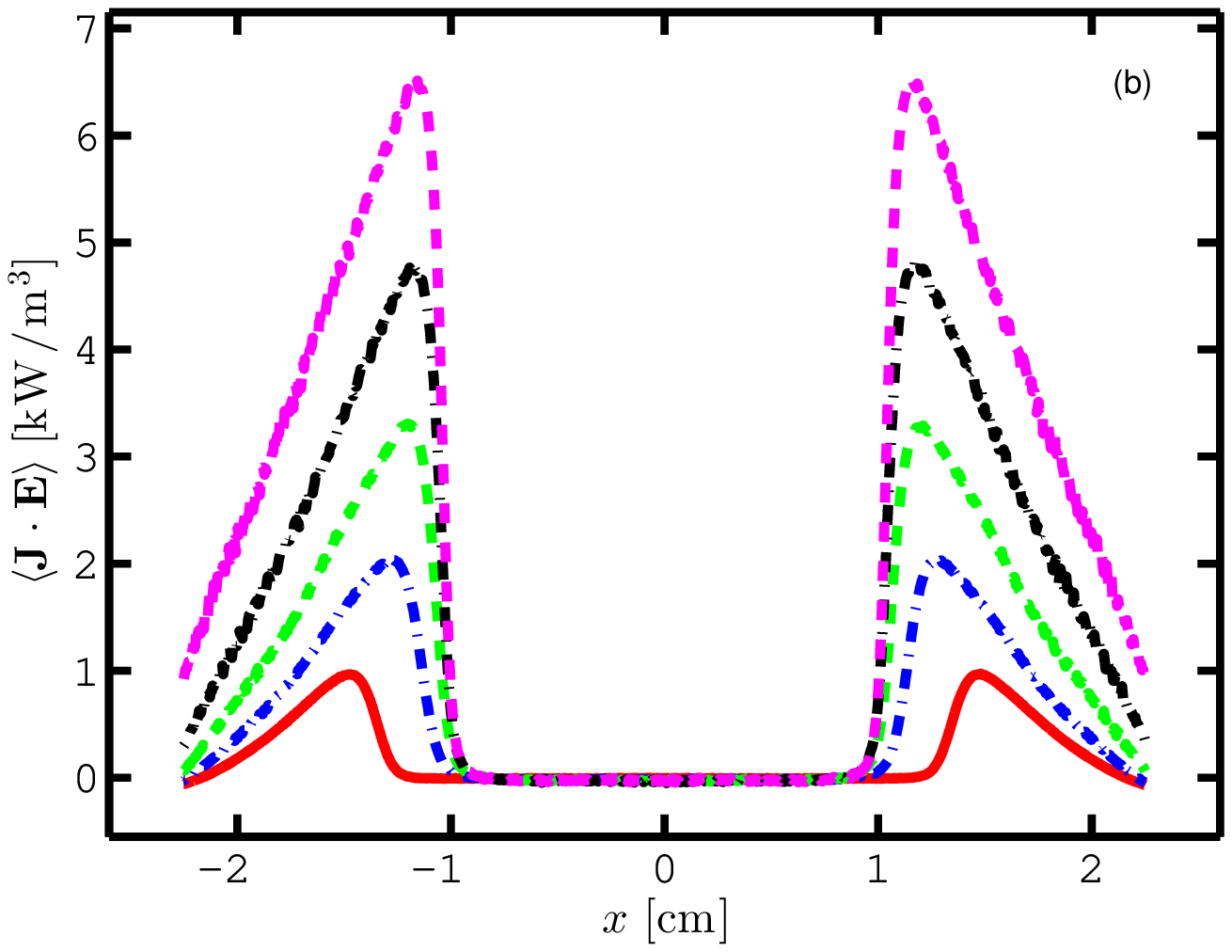}}
\caption{The time averaged electron heating profile for various voltage amplitudes for a parallel 
plate capacitively coupled oxygen discharge with a gap
separation of 4.5 cm driven at 13.56 MHz operated at (a) 10 mTorr, and (b)  50 mTorr.  }
\label{jdotave}       
\end{figure}

\begin{figure}
\resizebox{0.45\textwidth}{!}{%
  \includegraphics{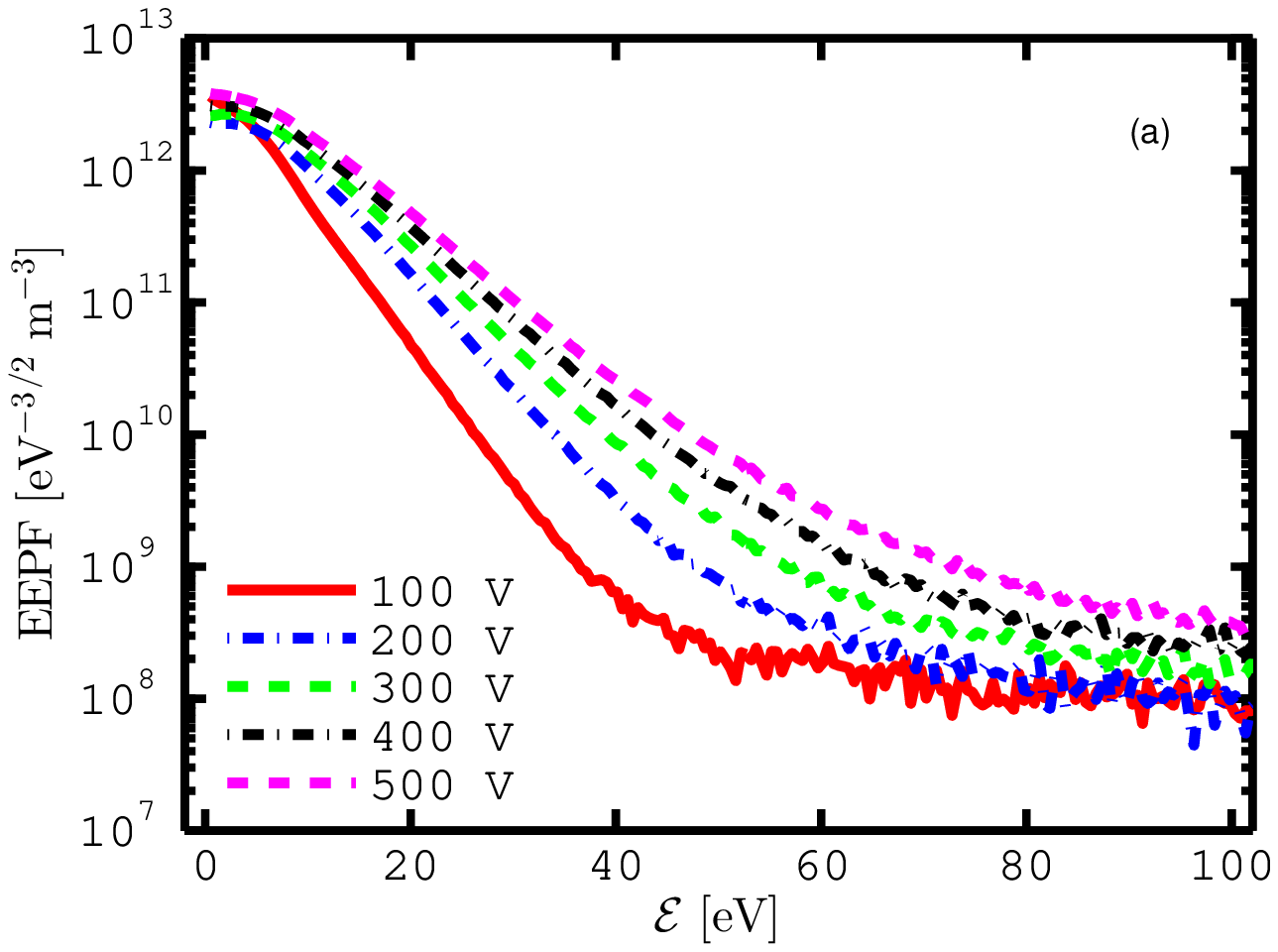}}
\resizebox{0.45\textwidth}{!}{%
  \includegraphics{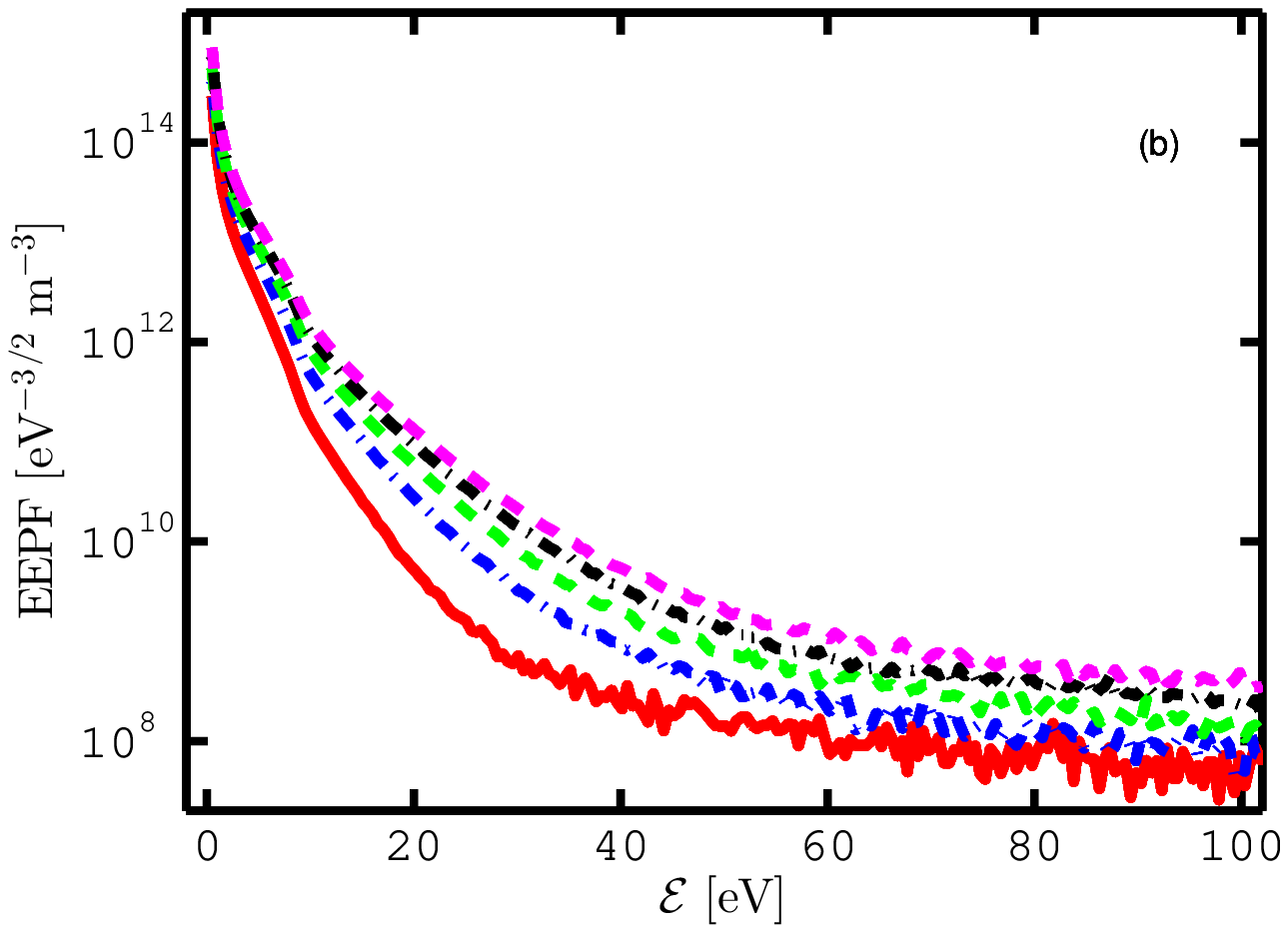}}
\caption{The electron energy probability function (EEPF) in the discharge center for  various voltage amplitudes for a parallel 
plate capacitively coupled oxygen discharge with a gap
separation of 4.5 cm driven at 13.56 MHz operated at (a) 10 mTorr, and (b)  50 mTorr.  }
\label{eepf}       
\end{figure}

\begin{figure}
\resizebox{0.45\textwidth}{!}{%
  \includegraphics{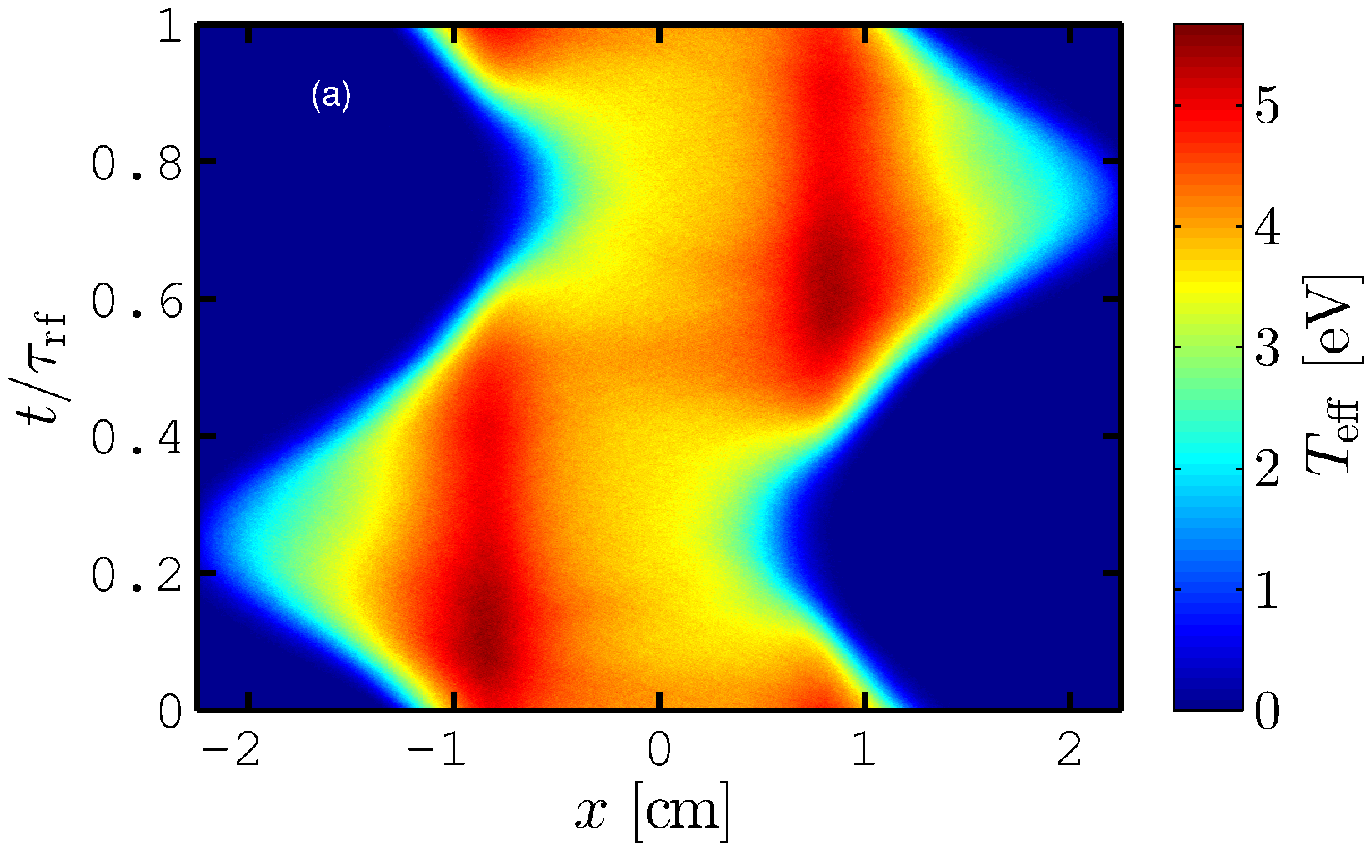}}
\resizebox{0.48\textwidth}{!}{%
  \includegraphics{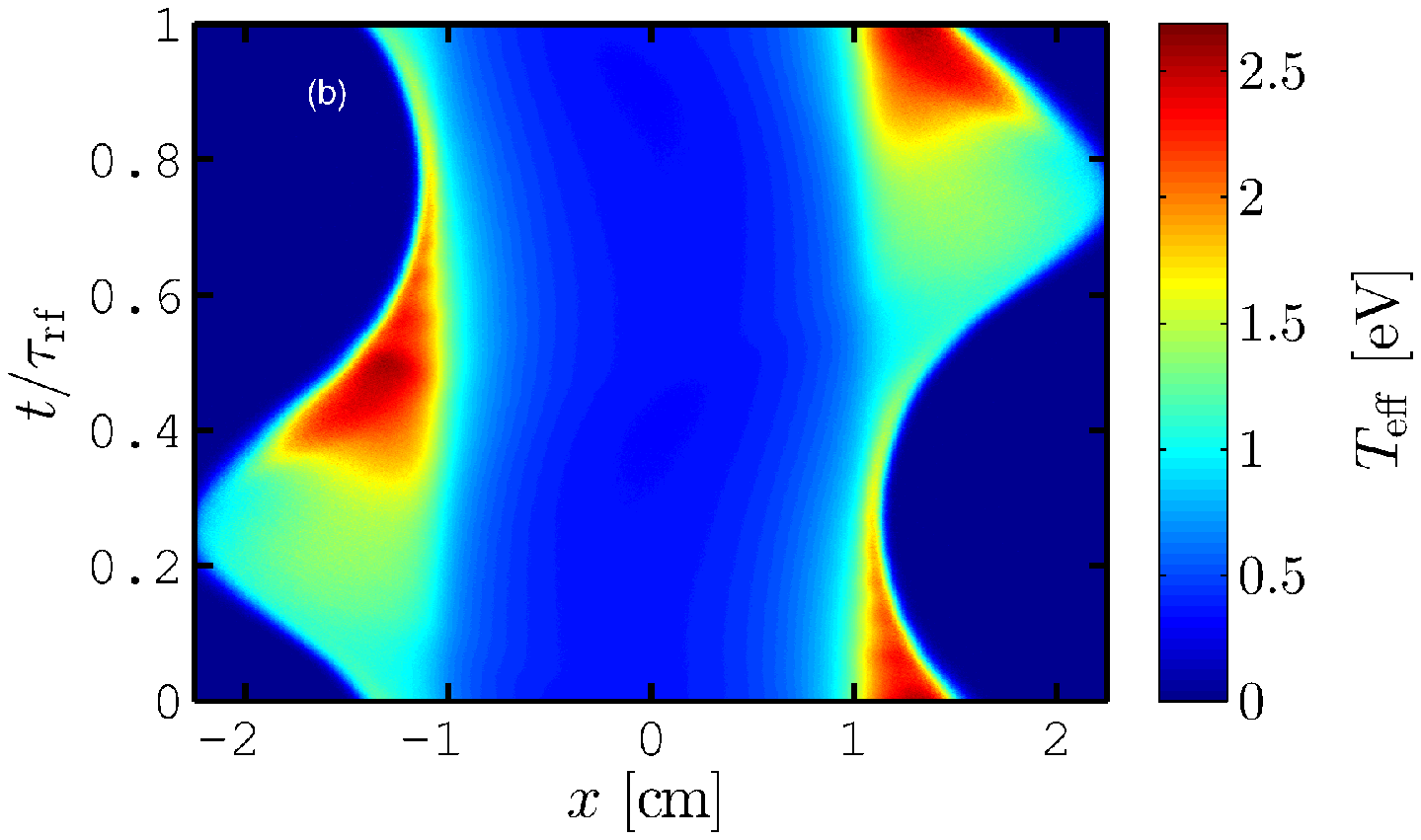}}
\caption{The spatio-temporal behavior of the effective electron temperature for a parallel 
plate capacitively coupled oxygen discharge with a gap
separation of 4.5 cm driven at 13.56 MHz  with $V_0 = 300$ V operated at (a) 10 mTorr, and (b)  50 mTorr.  }
\label{teff}       
\end{figure}

\begin{figure}
\resizebox{0.45\textwidth}{!}{%
  \includegraphics{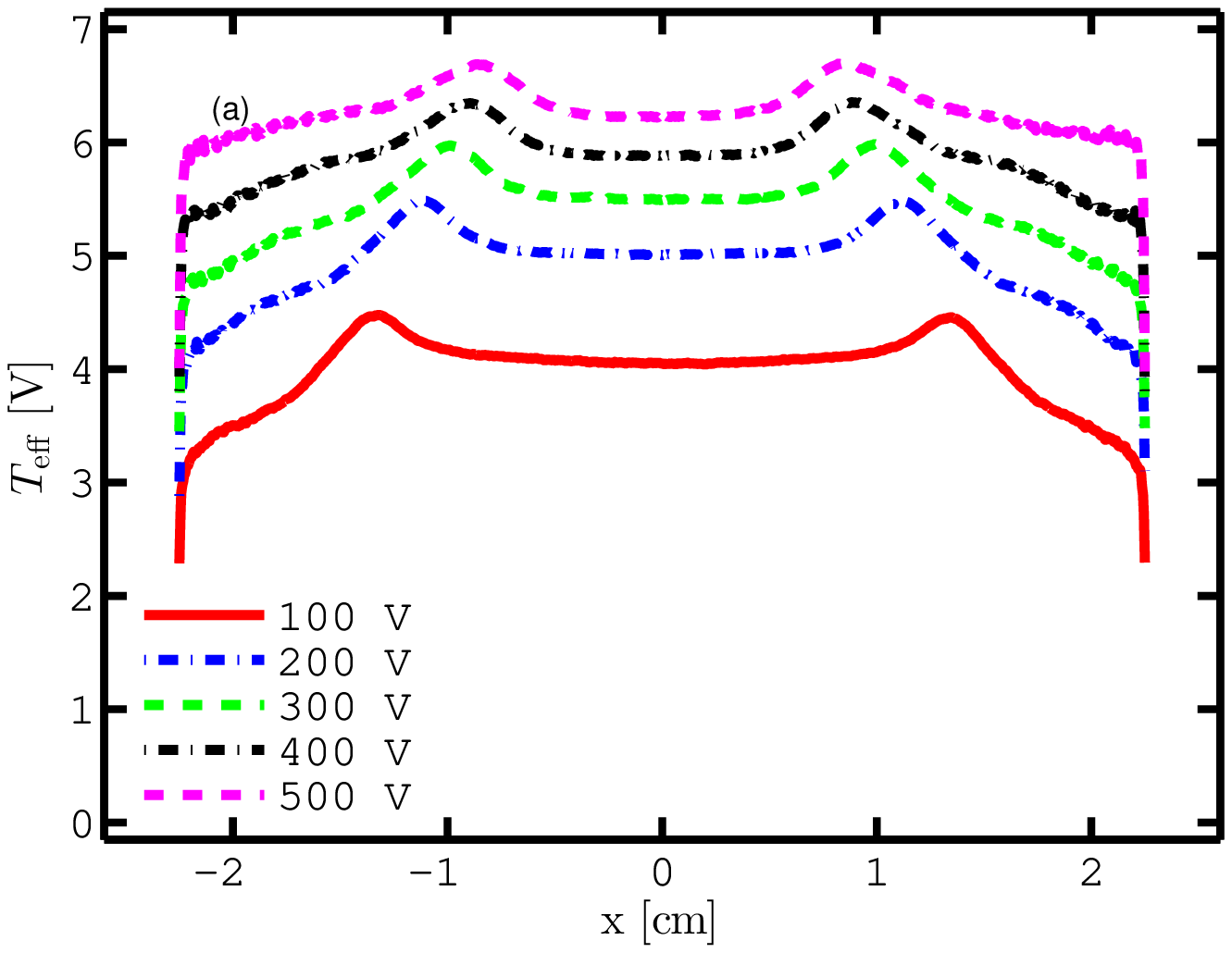}}
\resizebox{0.45\textwidth}{!}{%
  \includegraphics{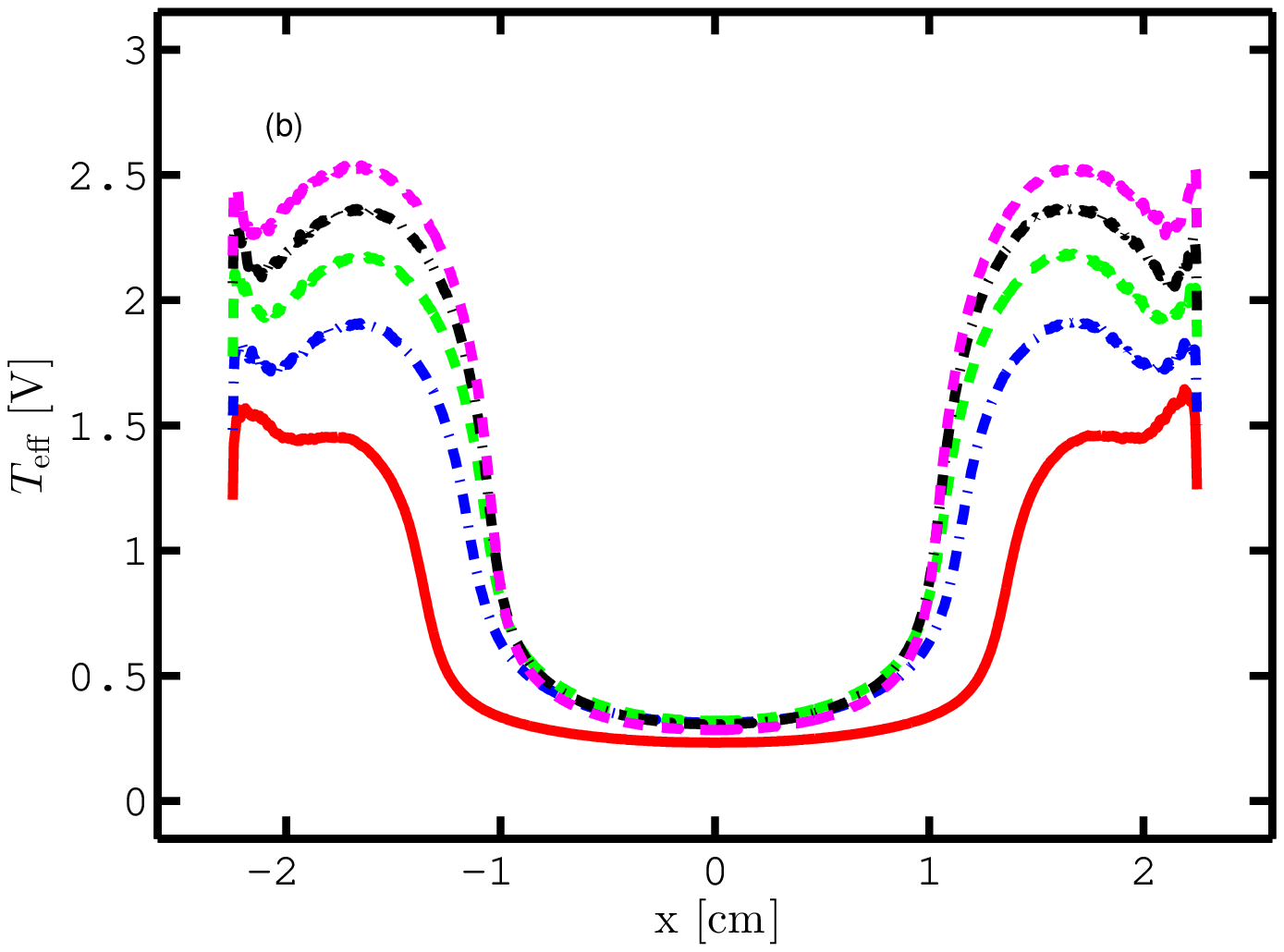}}
\caption{The time averaged effective  electron temperature for various voltage amplitudes for a parallel 
plate capacitively coupled oxygen discharge with a gap separation of 4.5 cm driven at 13.56 MHz operated at (a) 10 mTorr, and (b)  50 mTorr.  }
\label{teffave}   
\end{figure}

Figure  \ref{jdote}  shows the  spatio-temporal behavior of the electron heating 
rate  $ {\bf J}_{\rm e} \cdot {\bf E}$, 
where ${\bf J}_{\rm e}$ and ${\bf E}$ are the spatially and temporally varying electron current
density and electric field, respectively, for a discharge operated at 
 10 mTorr and 50 mTorr for voltage amplitude $V_0 = 300$ V. 
For each of the figures the abscissa covers the whole inter-electrode gap, from the powered electrode
on the left hand side to the grounded electrode on the right hand side. Similarly the ordinate covers 
the full rf cycle. Note that the color scale differs in magnitude between the two figures.  
Figure \ref{jdote} (a) shows the spatio-temporal behavior of the electron heating rate when operating at 10 mTorr
and Figure \ref{jdote} (b) shows the heating when operating at 50 mTorr.  
For both pressures the electron heating is most significant during the sheath expansion phase at 
each electrode (the red areas).  We also note that cooling in the sheath region 
during the sheath collapse is always apparent (note the different scale). 
At 10 mTorr a significant  energy gain (red and yellow areas) 
and small energy loss (dark blue areas) are also evident in the plasma bulk region as seen in Figure 
\ref{jdote} (a). The electron heating appears  during the sheath collapse on the bulk side of the sheath edge 
while there is cooling (electrons loose energy) on the electrode side (the lower left hand corner 
and upper center on the right hand side).
At 50 mTorr the electron heating rate in the sheath region 
has increased and there is almost no electron heating in the plasma bulk as seen in Figure \ref{jdote} (b).  
We note that there are  high frequency oscillations in the electron heating rate at both pressures
 adjacent to the expanding sheath edge similar to
 those reported by Vender and Boswell \cite{vender92:1331}. 
These are due the generation of an energetic electron beam during sheath
expansion that in turn can trigger a beam-plasma  instability at the electron plasma frequency.
Remainants of  excess  negative charges from the sheath collapse leads to a build up of an 
electric field that is large enough to accelerate bulk electrons toward the powered electrode.  
As the rf sheath expands again,  the electrons  are
accelerated back into the bulk plasma  with high kinetic energy.
This leads to an electron-electron two-stream  instability between the bulk electrons
and the electrons accelerated by the moving sheath  that is the cause of the oscillations  
\cite{oconnell07:034505}. These oscillations were first predicted  computationally 
 \cite{vender92:1331,meige08:1384} but have been more recently  been confirmed experimentally
using phase resolved optical emission spectroscopy (PROES)
\cite{oconnell07:034505,oconnel08:1382}.   The origins of the electric fields 
and the kinetics of multiple electron beams and the interactions of cold and hot electrons
 have been explored further more recently by \citet{wilczek16:063514}.
Figure \ref{jdotave} shows the time averaged electron heating rate profile  $\langle {\bf J}_{\rm e} \cdot {\bf E} \rangle$
 at 10 and 50 mTorr.  
We see in Figure \ref{jdotave} (a) 
that when the discharge is operated at 10 mTorr electron heating in the  electronegative
 core dominates.   
 This is in agreement with our earlier findings that at low pressures electron heating 
within the electronegative core dominates and 
the presence of the metastable molecules
has only a minor influence on the heating mechanism \cite{gudmundsson15:153302,gudmundsson17:120001}.  
We see in  Figure \ref{jdotave} (a) that the time averaged heating rate in the electronegative 
core increases with increased voltage amplitude. 
There is both electron heating and electron cooling apparent in the sheath regions.
At $V_0 = 100$ V the cooling cancels out the heating in the sheath region. With an increase 
in the voltage amplitude we see that the time averaged heating rate as well as the cooling rate 
in the sheath region increases.   When the discharge is operated at 
50 mTorr the time averaged  electron heating rate in the electronegative core is roughly zero, and 
the time averaged electron heating is almost entirely located in the sheath regions.  We also see that 
the time averaged electron heating rate in the sheath  region increases with increased voltage amplitude.  
We have earlier demonstrated how adding the singlet metastable molecules to the reaction set
drives the electron heating in the electronegative core to zero such that all the 
electron heating occurs only in the sheath regions  at operating pressures of 50 mTorr and higher 
\cite{gudmundsson15:035016,gudmundsson15:153302,gudmundsson17:120001}. 
We see that the sheath width increases with increased voltage amplitude at both 10 and 50 mTorr.  
The different heating mechanisms at 10 and 50 mTorr results in different shape of the 
electron energy distribution.  This can be seen in Figure 
\ref{eepf} that shows the electron energy probability function (EEPF) for discharges operated at 10 and 50 mTorr. 
We see that when the discharge is operated at 10 mTorr the EEPF is concave as seen in Figure \ref{eepf} (a) while 
at 50 mTorr it is convex or bi-Maxwellian, characterized by the two distinct low and high 
energy electron groups.  It is well known that when sheath heating or 
stochastic heating dominates in the capacitively 
coupled discharges the electron energy distribution can be described as a bi-Maxwellian.  
We see that at both 10 and 50 mTorr increasing the voltage amplitude increases the number 
of higher energy electrons. In all cases we also note a high energy tail, high energy electrons that are due 
to the secondary electron emission.  
These secondary electrons are emitted from the electrodes and are accelerated within the sheath 
and cause ionization as they travel through the plasma. 
The spatio-temporal behavior of the effective electron temperature ($T_{\rm eff} = (2/3) \langle {\cal E} \rangle$  
where  $\langle {\cal E}  \rangle$ is the  average electron energy)  is shown in Figure \ref{teff} for voltage amplitude 
of $V_0 = 300$ V.  It shows the effective electron temperature as a function of position between the 
electrodes within one rf period. At 10 mTorr seen in  Figure \ref{teff} (a)   we note  that the effective electron 
temperature    is
high within the plasma bulk (the electronegative core) throughout the rf period.  Furthermore, we see that  
the effective electron temperature peaks within the plasma bulk during the sheath collapse phase.  
At 50 mTorr, seen in  Figure \ref{teff} (b),  we see a  peak in the effective electron 
temperature  within the plasma bulk in the sheath expansion phase. 
At 50 mTorr the effective electron temperature is low within the plasma bulk
 throughout the rf period.  
The profiles of the time averaged effective electron temperature are shown in Figure \ref{teffave}. 
The time-averaged 
effective electron temperature   profile changes significantly when  the pressure is varied as seen by
comparing Figure \ref{teffave} (a) for 10 mTorr and  Figure \ref{teffave} (b) for 50 mTorr.  
 When the  discharge is operated at 10 mTorr the electron temperature is high and increases with increased voltage 
amplitude as seen in Figure \ref{teffave} (a). For $V_0 = 100 $ V the effective electron temperature 
in the discharge center is $T_{\rm eff} = 4.1$ eV and 
for  $V_0 = 500 $ V it is $T_{\rm eff} = 5.9$ eV.
 When operated at 50 mTorr the electron temperature is low, highest in the 
sheath region and drops to roughly 0.2 -- 0.3 eV within the electronegative core.  
In the sheath region the effective electron temperature 
increases with increased voltage amplitude from roughly 1.5 eV when $V_0 = 100$ V to roughly 2.5 V  when $V_0 = 500$ V.
These results are consistent with the Langmuir probe measurements of the effective electron temperature 
repoted by Kechkar et al.~in a slightly  geometrically asymmetric 
capacitively coupled oxygen discharge with electrodes made of aluminum alloy 
\cite{kechkar15t,kechkar17:065009},  where the driven electrode was 
205 mm in diameter and the grounded electrode was 295 mm in diameter  
with electrode separation of 45 mm. In the discharge center
at 10 mTorr and 200 W ($V_0 = 338$ V)  they find $T_{\rm eff} = 4.5$ eV and in the pressure range 50 -- 100 mTorr,  
$T_{\rm eff} \approx 0.7$ eV ($V_0 = 290$ V at 50 mTorr) \cite{kechkar15t,kechkar17:065009,kechkar16}.  

We would expect collisionless (stochastic)  heating at low pressures when the 
electron-neutral mean free path $\lambda_{\rm en}$, is comparable or greater than the gap between the electrodes
$L$ or the width of the plasma bulk $L_{\rm bulk}$
or $\lambda_{\rm en} > L_{\rm bulk}$.  At high pressures the electron-neutral mean free path $\lambda_{\rm en}$ is 
small so that electrons
collide more frequently with the neutral background gas or $\lambda_{\rm en} <  L_{\rm bulk}$.   
The  electron-neutral mean free path is $\lambda_{\rm en} \approx 50$ mm at 10 mTorr for 
effective electron temperature in the range 4 -- 6 eV.  These electrons have a mean free 
path that is much larger than the width of the plasma bulk $L_{\rm bulk} \approx 20$ mm.
 At 50 mTorr  $\lambda_{\rm en} \approx 12$ mm for 
effective electron temperature of 0.5 eV and $\lambda_{\rm en} \approx 9$ mm for 
effective electron temperature of 2 eV.  Thus electron neutral collisions are
rare in these discharges. For the secondary electrons   $\lambda_{\rm en} \approx 217$ mm at 10 mTorr
and  $\lambda_{\rm en} \approx 43$ mm at 50 mTorr if we assume acceleration up to 100 eV, and these values 
increase with increased acceleration voltage.  
At both 10  and 50 mTorr these high energy electrons have a mean free 
path that is larger than the width of the plasma bulk. 
Thus we would not expect the secondary electrons to have much influence at 10 and 50 mTorr and to be
mostly lost to the electrodes without collisions with the neutral molecules.
These calculations are based on the momentum transfer cross section given by Itikawa \cite{itikawa09:1}.
Electron-neutral collisions are thus not a very efficient heating mechanism at these pressures so something more
has to come to play, in order to explain the observed electron heating within the plasma bulk at 10 mTorr.

\begin{figure}
\resizebox{0.45\textwidth}{!}{%
  \includegraphics{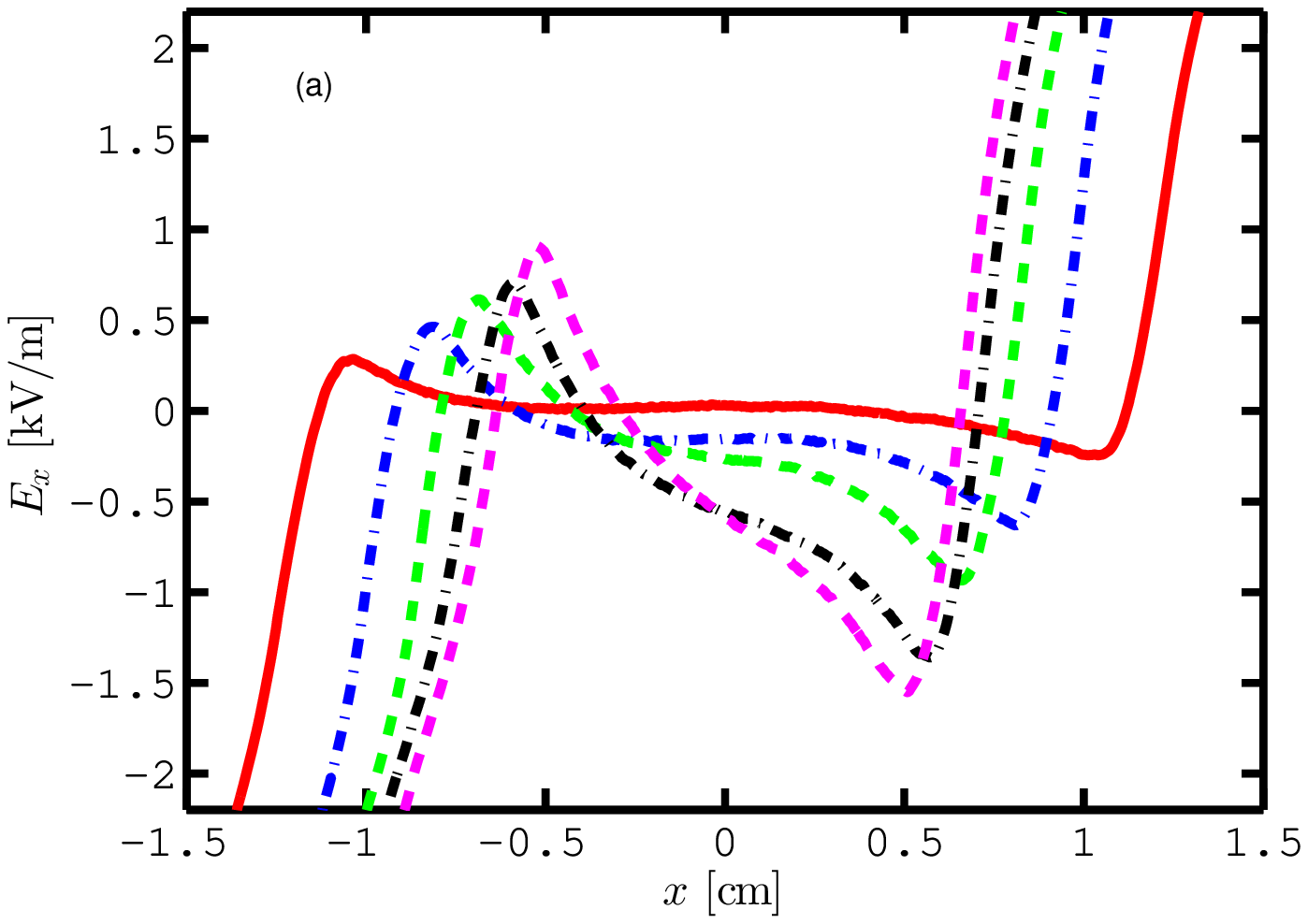}}
\resizebox{0.45\textwidth}{!}{%
  \includegraphics{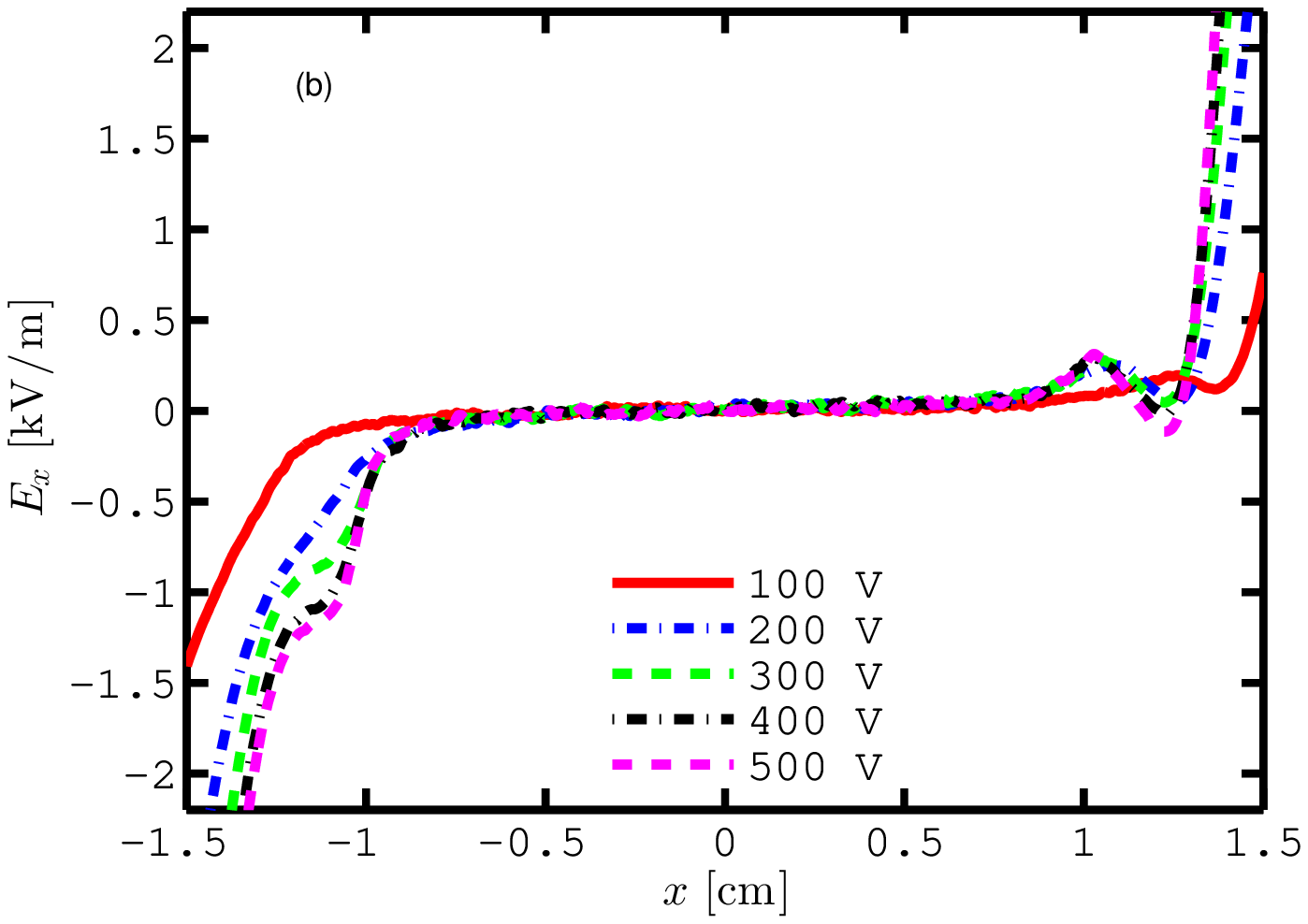}}
\caption{The axial electric field for a parallel 
plate capacitively coupled oxygen discharge with a gap
separation of 4.5 cm driven at 13.56 MHz operated at $t/\tau_{\rm rf} = 0.5$ at (a) 10 mTorr, and (b)  50 mTorr.  }
\label{exfield}       
\end{figure}

Figure \ref{exfield} shows the axial electric field at $t/\tau_{\rm rf} = 0.5$ for both 10 and 50 mTorr. 
At 10 mTorr (Figure \ref{exfield} (a)) we see that there is a significant electric field 
strength within the electronegative core.
The  electric field strength  and its gradient increase with increased voltage amplitude. 
This strong electric field within the plasma bulk (the electronegative core) indicates a
drift-ambipolar heating mode \cite{schulze11:275001}.    This electric field is a combination of a 
drift field and an ambipolar field.  The ambipolar field  is due to local maxima of the electron 
density at the sheath edge and a steep electron density gradient and 
yields the local maxima in the electric field observed at the sheath edges  \cite{schulze11:275001}.  
The drift electric field is due to  low bulk conductivity,
as we see later the electron density in the electronegative core is indeed very low.  This high electric field
accelerates the electrons to high average energies and thus causes ionizations within the plasma bulk. 
Thus when the discharge is operated at 10 mTorr the electron heating 
consists of stochastic heating in the sheath region and DA-heating within the electronegative core. 
We saw in Figure \ref{jdote} (a) that the electron heating maxima occurs within the plasma bulk
and close to the collapsing sheath edge.  The DA-mode is characterized by a high ionization rate 
and high electron energy within the plasma bulk. 
Thus we can say that at 10 mTorr the discharge is operated in a combined DA- and $\alpha$-mode.
 The DA-heating mode has been observed
in  electronegative discharges, including  a dual frequency 
oxygen discharge \cite{liu12:114101}, single frequency silan discharge \cite{yan00:3628} 
and CF$_4$ discharge \cite{schulze11:275001,liu15:034006}, but also in dusty plasmas 
including single frequency argon discharge \citep{killer13:083704} and hydrogen diluted 
silane discharge \citep{schungel13:175205}. 
Furthermore, ohmic heating (drift) mode ($\Omega$ mode) has been observed in  an atmospheric-pressure
diffuse dielectric barrier discharge in helium \citep{boisvert17:035004}. 
It is referred to as the   DA mode when there is a 
simultaneous  presence  of  both  this ohmic  heating 
($\Omega$ mode)  and  the  heating  due  to  the  ambipolar  field  in 
electronegative discharges.
When the discharge is operated at 50 mTorr  (seen in Figure \ref{exfield} (b)) the electric field is
zero within the electronegative core. We saw in Figure \ref{jdote} (b) that electron heating is almost
solely in the sheat regions at 50 mTorr.  Hence, at 50 mTorr the discharge is operated in a pure $\alpha$-mode. 
The transition from combined DA-$\alpha$-mode to pure $\alpha$-mode coincides with a significant  
decrease in the electronegativity as discussed below.  
Transitions  between  the DA-mode and the $\alpha$-mode have been demonstrated by both simulations 
and experiments on  CF$_4$ discharges \citep{schulze11:275001,liu15:034006}. 
By increasing the pressure at a fixed 
voltage,  a  transition  from  the  $\alpha$-mode  to  the  DA-mode  is 
induced.  Note that the CF$_4$ discharge is weakly electronegative
at 75 mTorr while it is strongly electronegative at 600 mTorr \citep{derzsi15:346}.  
Also by increasing  the  voltage  at  a fixed  pressure,  a transition from the DA-mode 
to the $\alpha$-mode is observed in a CF$_4$ discharge \citep{schulze11:275001}.
  Here we show the opposite, that by increasing the pressure at a given voltage 
a transition from the DA-$\alpha$-mode to the $\alpha$-mode is observed in the oxygen discharge. 
This is a similar transition as reported by  Derzsi et al.~\cite{derzsi17:034002} 
which observe an  operation mode transition from DA-$\alpha$-mode to $\alpha$-mode in an oxygen discharge 
as harmonics are added to the voltage waveforms for 10 and 15 MHz driving frequency,
which also  coincides with a strong decrease in the electronegativity.  Earlier we have demonstrated a 
transition from the DA-$\alpha$-mode to pure $\alpha$-mode, for a discharge operated at 
 50 mTorr, when the singlet metastables were added to the  reaction set \cite{gudmundsson17:120001}.

\begin{figure}
\resizebox{0.45\textwidth}{!}{%
  \includegraphics{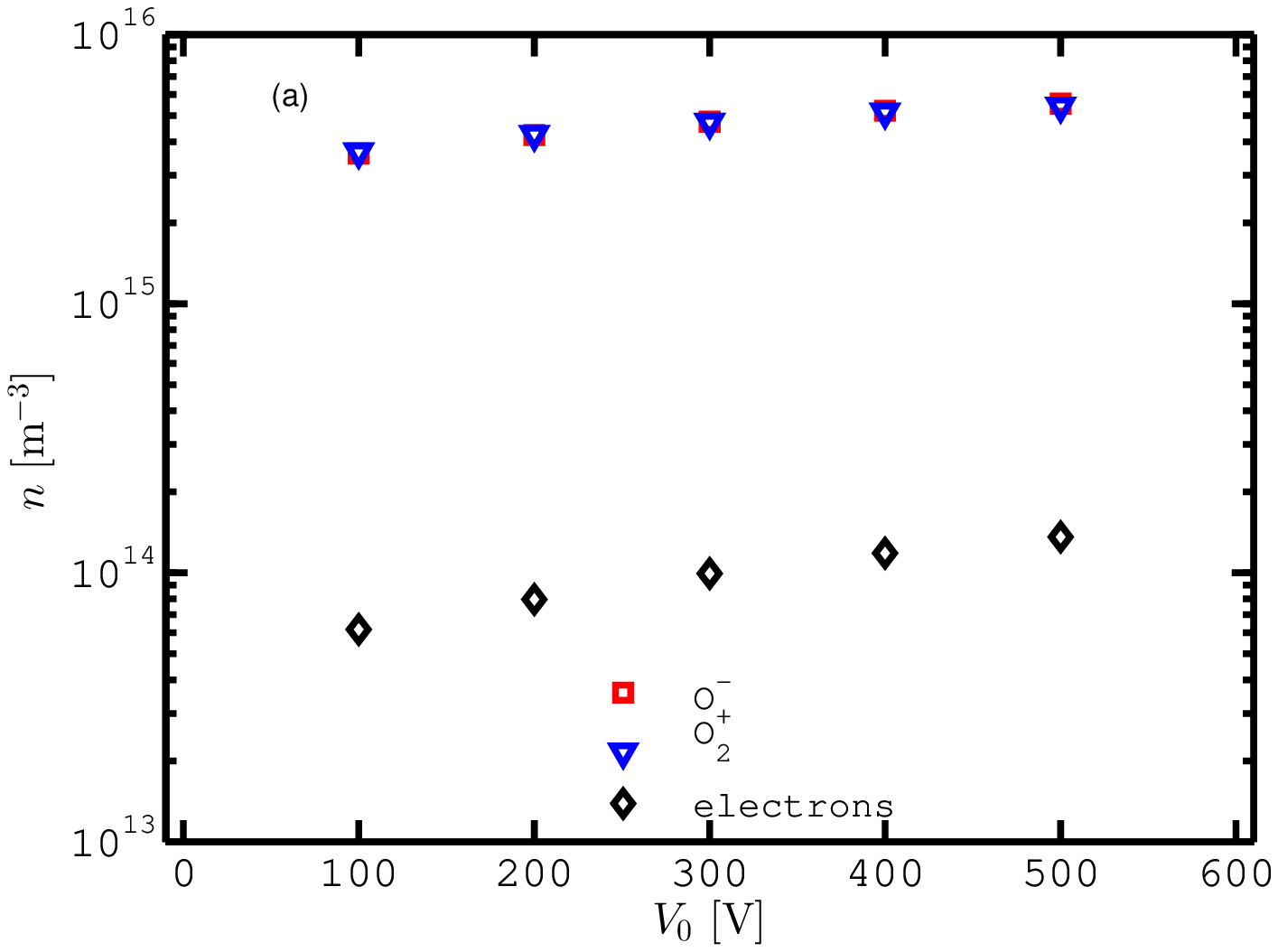}}
\resizebox{0.45\textwidth}{!}{%
  \includegraphics{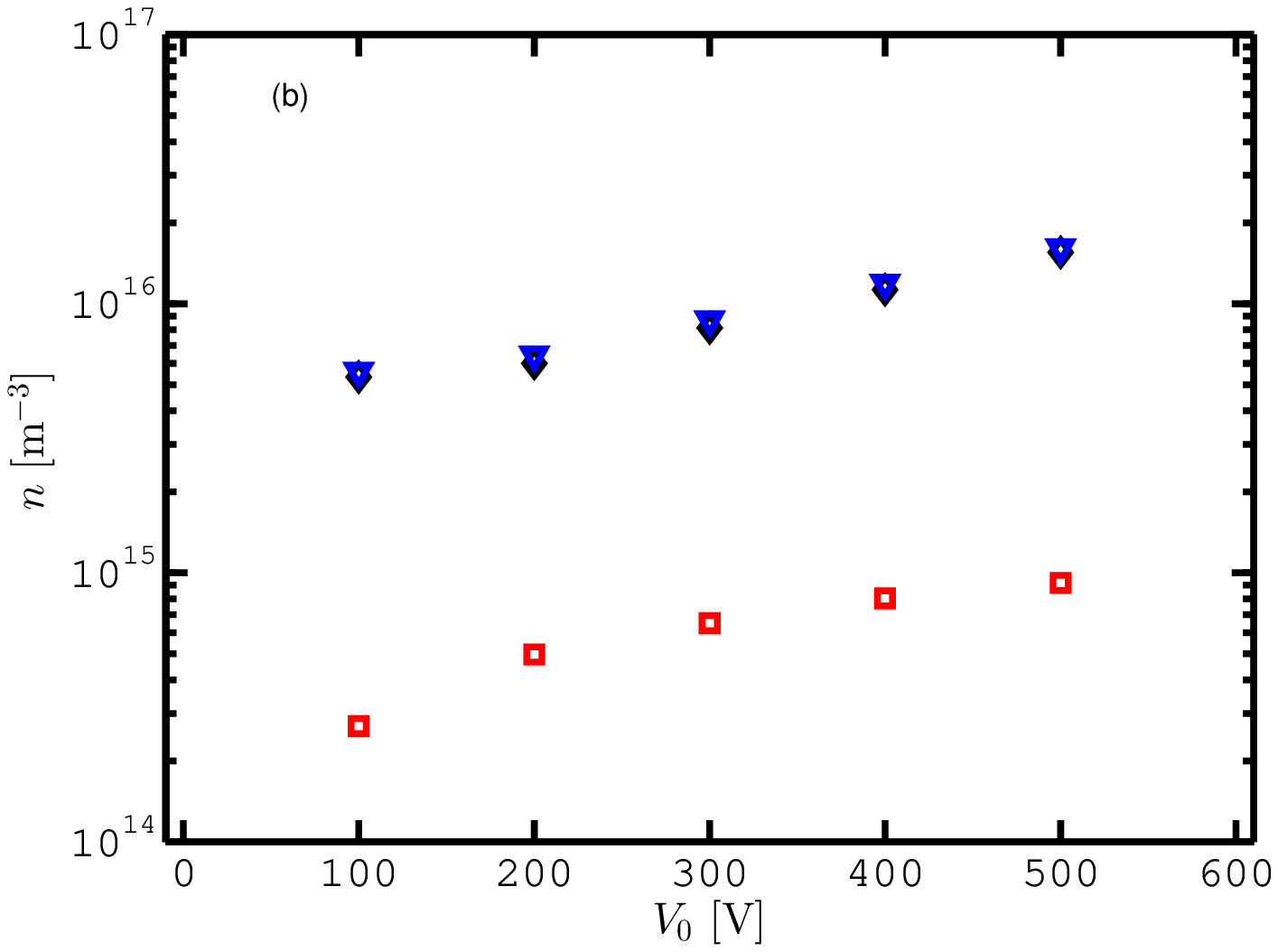}}
\caption{The center charged particle density versus the voltage amplitude for a parallel 
plate capacitively coupled oxygen discharge with a gap
separation of 4.5 cm driven at 13.56 MHz operated  at (a) 10 mTorr, and (b)  50 mTorr.  }
\label{centerdens}       
\end{figure}

\begin{figure}
\resizebox{0.45\textwidth}{!}{%
  \includegraphics{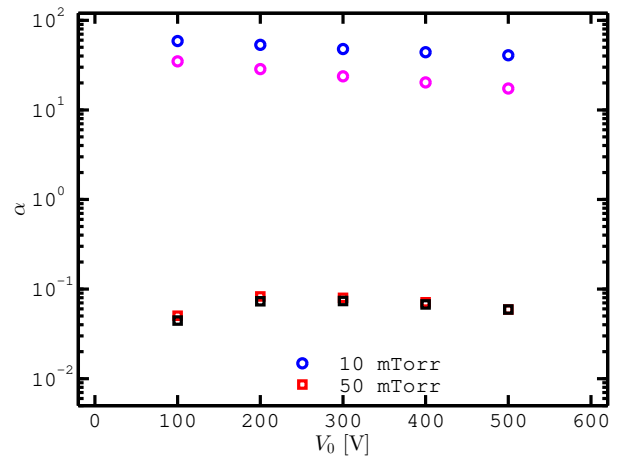}}
\caption{The center electronegativity $\alpha_0$ and the 
average electronegativity $\alpha_{\rm ave}$ versus the voltage amplitude for a parallel 
plate capacitively coupled oxygen discharge operated  at $\circ$ 10 and $\boxempty$ 50 mTorr with a gap
separation of 4.5 cm driven at 13.56 MHz. }
\label{alpha}       
\end{figure}

The center density of the dominating charged particles is shown in Figure \ref{centerdens}.  When operating at 
10 mTorr the O$_2^+$-ion density and the O$^-$-ion density are similar and the electron density is significantly smaller
as seen in Figure \ref{centerdens} (a). The electron density $n_{\rm e}$ is only $0.6 \times 10^{14}$ m$^{-3}$ 
at $V_0 = 100$ V and increases to $1.3 \times 10^{14}$  m$^{-3}$  at $V_0 =  500$ V.  Thus the conductivity $\sigma_{\rm dc} \propto n_{\rm e}$ 
is low in the plasma bulk.      
 When operating at 50 mTorr the O$_2^+$-ion density and the electron  
density are similar and the O$^-$-ion density is significantly smaller
as seen in Figure \ref{centerdens} (b). At 50 mTorr the electron density is much higher than at 10 mTorr and 
increases from  $0.5 \times 10^{16}$ m$^{-3}$  at $V_0 =  100$  V and  $1.6 \times 10^{16}$  m$^{-3}$  at $V_0 =  500$ V. 
For comparison  Kechkar et al.~\cite{kechkar15t,kechkar17:065009,kechkar16} measured 
the electron density in  a slightly asymmetric 
capacitively coupled oxygen discharge to be $6.5 \times 10^{14}$ at  30 W ($V_0 = 90$ V)
and  $2.7 \times 10^{15}$ at  200 W ($V_0 = 338$ V) when operated at 10 mTorr,  
and $1.6 \times 10^{15}$ at   30 W ($V_0 = 85$ V)  and  $3 \times 10^{15}$ at   200 W ($V_0 = 290$ V) when operated at 50 mTorr.
As seen by comparing Figures \ref{centerdens} (a) and (b) the 
negative ion density is higher at 10 mTorr than at 50 mTorr.  At 10 mTorr and 100 V the 
O$^-$-ion density is  $3.6 \times 10^{15}$ m$^{-3}$   at 100 V and  increases to $5.5  \times 10^{15}$ m$^{-3}$ at 500 V.  
At 50 mTorr  the  O$^-$-ion density is  $2.7 \times 10^{14}$ m$^{-3}$   at 100 V and  increases to $9.2 \times 10^{14}$ m$^{-3}$ at 500 V.
So as the pressure is increased the electron density increases and the negative ion density decreases so that the 
electronegativity decreases significantly.  We know from global (volume averaged) model studies of the
oxygen discharge that at low pressure ( < 10 mTorr) the destruction of negative ions is dominated by 
electron impact detachment while at higher pressure detachment by  oxygen atoms and singlet metastable oxygen molecules and charge exchange with the ground state molecule  take over  and their role increases with increased discharge pressure
and the negative ion density decreases as the discharge pressure is increased \citep{toneli15:325202}. 
 At 10 mTorr the negative O$^-$-ions are effectively created by the dissociative attachment processes as the 
effective electron temperature is high, and the detachment by electrons is dominating.  
At 50 mTorr due to the low effective electron temperature dissociative 
attachment is not very effective in creating the negative
O$^-$-ions while they are effectively eliminated by the detachment processes which are roughly independent of the 
electron temperature.
The center electronegativity $\alpha_0 = n_{-0}/n_{\rm e0}$, where $n_{-0}$ is the center negative ion 
density and $n_{{\rm e}0}$ is the center electron density, is shown versus the voltage amplitude in 
Figure \ref{alpha}. The electronegativity is significantly higher when operating at 10 mTorr than
when operating at 50 mTorr.  We see that the electronegativity at 10 mTorr 
decreases with increased voltage amplitude
 from 58 for $V_0 = 100$ V to 40  for $V_0 = 500$ V.   At 50 mTorr the electronegativity is 0.05 for
$V_0 = 100$ V, 0.08 for $V_0 = 300$ V, and 0.06 for $V_0 = 500$ V.
Figure \ref{alpha} also shows the average  electronegativity 
\begin{equation}
\alpha_{\rm ave} = \frac{\int_0^L n_-(x) dx}{\int_0^L n_{\rm e}(x) dx}
\end{equation}
versus the voltage amplitude, where $n_-$ is the O$^-$-ion density, $n_{\rm e}$ 
is the electron density and $L$ is
the electrode separation.   We see that the average electronegativity 
is somewhat lower than the center electronegativity 
at 10 mTorr while at 50 mTorr the two are very similar.  Experimentally, 
 \citet{katsch00:323} estimated the electronegativity 
in the discharge center of a
capacitively coupled oxygen discharge to be roughly 2 at
103 mTorr and 150 V and to fall below unity as the applied voltage was increased to 280 V. 
Also, \citet{berezhnoj00:800} determined the electronegativity in a capacitively coupled oxygen discharge to 
be roughly 10 in the pressure range 22.5 -- 225 mTor.

\section{Conclusion}
\label{conclusion}

The one-dimensional object-oriented particle-in-cell Monte Carlo 
collision code {\tt oopd1}
was applied  to explore the  evolution of the electron heating 
mechanism, the EEPF, and the effective electron temperature, 
in  a capacitively coupled oxygen discharge with the applied voltage. 
We compare operation at 10 mTorr and 50 mTorr.  We demonstrate that there is a significant difference,
the electron heating mechanism is different,  which leads to very different electron energy probability 
function and then very different time averaged  electron temperature profile for the two different operating pressures.
There is a significant electron heating in the electronegative core and high effective electron temperature 
that increases with increased applied voltage when operating at 10 mTorr. At 50 mTorr the effective 
 electron temperature is very low (roughly 0.2 -- 0.3 eV) in the electronegative core at all voltages. 
Furthermore, there is significant difference in electronegativity.
We observe a strong electric field within the plasma bulk when operating at 10 mTorr while 
the electric field is zero within the plasma bulk when operating at 50 mTorr.
 At 10 mTorr the discharge is operated in combined drift-ambipolar and $\alpha$-mode
and at 50 mTorr it is operated in a pure $\alpha$-mode.

\acknowledgements

This work was partially supported by the  Icelandic Research Fund Grant
No.~163086,  the University of Iceland Research Fund, and 
the Swedish Government Agency for Innovation Systems (VINNOVA) contract no. 2014-04876.
 The authors acknowledge discussions with H{\'o}lmfr{\'i}{\dh}ur Hannesd{\'o}ttir
who also developed the capability of plotting the spatio-temporal figures.



%
%

%
%

\end{document}